\documentclass[twocolumn,
reprint,
showpacs,
preprintnumbers,
aps,
pra,
letterpaper,
superscriptaddress,
nofootinbib,
tightenlines,
floats,
floatfix]{revtex4}
\usepackage{graphicx}
\usepackage{dcolumn}
\usepackage{bm}
\usepackage{latexsym}
\usepackage{amsmath,amssymb}
\usepackage[draft=false]{hyperref}
\usepackage[latin1]{inputenc}
\usepackage{latexsym}
\usepackage{amsmath}
\usepackage{ifpdf}

\begin{document}


\title{Space-Time Curvature Signatures in Bose-Einstein Condensates}

\author{Tonatiuh Matos\footnote{Part of the Instituto Avanzado de Cosmolog\'ia 
(IAC) collaboration http://www.iac.edu.mx/}}
 \email{tmatos@fis.cinvestav.mx}\affiliation{Departamento de F\'{\i}sica,
 Centro de Investigaci\'on y Estudios Avanzados del IPN\\
 A. P. 14--740,  07000, M\'exico, D.F., M\'exico.}
\author{Eduardo Gomez}
 \email{egomez@ifisica.uaslp.mx}\affiliation{Instituto de F\'isica, Universidad Aut\'onoma de San Luis Potos\'i, San Luis Potos\'i 78290, M\'exico}


\begin{abstract}
We derive a generalized Gross-Pitaevski (GP) equation immersed on a electromagnetic and a weak gravitational field starting from the covariant Complex Klein-Gordon field in a curved space-time. We compare it with the GP equation where the gravitational field is added by hand as an external potential. We show that there is a small difference of order $g z/c^2$ between them that could be measured in the future using Bose-Einstein Condensates (BEC). This represents the next order correction to the Newtonian gravity in a curved space-time.  
\end{abstract}

\pacs{67.85.Jk, 37.25.+k, 04.80.-y}

\maketitle


Among the four forces of nature, the gravitational force is the hardest one to study.
It is orders of magnitude weaker than the other three forces and requires astronomical
masses to see corrections beyond the basic Newtonian formula. The lab scale masses
that we can manipulate introduce gravitational forces that are very hard to measure
accurately. This is reflected in the poor precision that we currently have on the
determination of the Newtonian constant of gravitation 
($G=6.67384(80) \times 10^{-11}$kg$^{-1}$m$^3$s$^{-2}$) \cite{Codata2010}.

Most of the experiments that measure the gravitational force use basically
the same method pioneered by Cavendish \cite{Mohr12}. Atomic interferometers
have been recently introduced as an alternative way to implement sensitive
gravimeters \cite{Kasevich91}. Here, cold atoms are launched vertically
and subjected to a sequence of laser pulses that split, recombine and interfere
the atoms. Using a Bose-Einstein Condensate (BEC) reduces the ballistic expansion
of the atomic cloud and allows for a larger fall time thus increasing the
precision of the gravimeters.

Sensitive atomic gravimeters are useful in the detection of gravitational
waves \cite{Dimopoulos08}, tests to Einstein's general relativity
\cite{Dimopoulos07}, study of short range gravitational forces \cite{Ferrari06}
and coherent evolution of delocalized quantum objects \cite{Lamine06}. 
Measurements of $g$ (the gravitational acceleration of the earth) typically
reach 9 digits of precision. This has been recently improved by two order of
magnitude by launching a BEC in free fall in a 10 m tower \cite{Dickerson13}.
Planned improvements to this include using a 146 m tower \cite{Zoest10}, 
sending the atoms in parabolic flights \cite{Geiger11} or putting the atoms 
in space \cite{Tino13}.

BECs are usually described by a Gross-Pitaevskii (GP) equation, where an external
field is added in order to confine the system or to include the 
earth's gravitational acceleration (\cite{Lemerzhal}).
Nevertheless, the GP equation is non-relativistic and non-covariant; it is 
not invariant under Lorenz transformations unlike the electromagnetic field. 
A GP equation obtained this way could only show Newtonian effects
of gravity. A more complete description is obtained by treating the
gravitational field not as a force, but as the curvature of space-time
as in Einstein's theory of relativity. In Ref. \cite{Suarez:2011yf} it was 
shown that the Klein-Gordon equation
in a curved space takes the GP form after a simple transformation and includes
extra terms that are interpreted as the gravitational field and a finite temperature
contribution. The gravitational components include terms beyond the Newtonian
gravity. In this work we propose that these terms could become the focus 
of future experimental measurements.

\section{Klein--Gordon in a weak gravitational field}

References \cite{Matos:2011kn} and \cite{Matos:2011pd} derive the relativistic
theory of a BEC in a quantum field theory description. Starting from the 
Klein-Gordon equation in a flat space-time, a generalized GP equation is obtained
for relativistic and finite temperature fields. The generalization for an 
expanding universe is given in \cite{Suarez:2011yf}. Here we apply the same 
approach on a curved space for a BEC living in a weak gravitational field, the Klein Gordon 
equation for a complex scalar field is given by
\begin{eqnarray}
\Box\Phi&-&\frac{dV}{d\Phi^*}=\nonumber\\
(\nabla^{\mu}+i\frac{e}{\hbar c}A^{\mu})(\nabla_{\mu}+i\frac{e}{\hbar c}A_{\mu})\Phi&-&\frac{dV}{d\Phi^*}=\frac{2m^2}{\hbar^2}U_{ext}{\Phi},\nonumber\\
\label{eq:KG}
\end{eqnarray}
where $\Phi$ is the complex scalar field and $A_{\mu}$ is the corresponding electromagnetic four vector.
In this work we use the Mexican hat scalar field potential given by
\begin{equation}
V=\frac{m^2c^2}{\hbar^2}\Phi\Phi^*
+\frac{\lambda}{4\hbar^2}|\Phi|^4.
\end{equation}
We add an external interaction $U_{ext}$ that can represent for example 
the potential that confines the condensate like a laser, a box or the gravitational field seen as an external one.
The metric of a weak gravitational field is \cite{Ma} 
\begin{equation}
ds^2=-(1+2\psi)dt^{2}+ (1-2\phi)g_{ij}dx^{i}dx^{j},
\end{equation}
where $\psi$, $\phi$ are the gravitational potentials and $g_{ij}$ is the 3-dimensional
flat-space metric. The Klein-Gordon equation with this metric reads
\begin{eqnarray}
\Box_{NEW}\Phi&-&\frac{dV}{d\Phi^*}+i\hat e\Phi\nabla_{\mu}A^{\mu}-\hat e^2A^2\Phi+2i\hat eA^{\mu}\nabla_{\mu}\Phi=\nonumber\\
\Box_{NEW}\Phi&-&\frac{dV}{d\Phi^*}
+i\hat e[(1+2\phi)\nabla\cdot{\textbf A}-\frac{1}{c}(1-2\psi)\dot\varphi\nonumber\\
&+&\nabla(\psi-\phi)\cdot{\textbf A}+\frac{1}{c}(\dot\psi+3\dot\phi)\varphi]\Phi\nonumber\\
&-&\hat e^2((1+2\phi)A^2-(1-2\psi)\varphi^2)\Phi\nonumber\\
&+&2i\hat e((1+2\phi){\textbf A}\cdot\nabla\Phi-\frac{1}{c}(1-2\psi)\varphi\dot\Phi)\nonumber\\
&=&\frac{2m^2}{\hbar^2}U_{ext}{\Phi}
\label{eq:KGWF}
\end{eqnarray}
with $\hat e=e/(\hbar c)$, ${\textbf A}$ and $\varphi$ the magnetic vector and electric scalar potentials such that $A_{\mu}=(\varphi,{\textbf A})$.
 The metric modifies the D'Alambertian as
\begin{eqnarray}
\Box_{NEW}&=&-(1-2\psi)\frac{\partial^2}{c^2\partial t^2}+(1+2\phi)\nabla^2\nonumber\\
&+&\frac{1}{c}[\dot\psi+3\dot\phi]\frac{\partial}{c\partial t}+\nabla(\psi-\phi)\cdot\nabla
\end{eqnarray}
The metric not only produces the gravitational potential, but it also affects the kinetic energy term, something that does not happen in the traditional treatment that adds the gravitational potential by hand.

In a curved space-time in the weak field limit the Maxwell equations read
\begin{eqnarray}
{\nabla}\cdot{\textbf E}+{\nabla}(\psi-3\phi)\cdot{\textbf E}&=&\rho\nonumber\\
\dot{\textbf E}+(\dot\psi-3\dot\phi){\textbf E}+{\textbf B}\times\nabla(\psi-3\phi)-\nabla\times{\textbf B}&=&{\textbf J}\nonumber\\
\nabla\cdot{\textbf B}-4{\textbf B}\cdot\nabla\phi&=&0\nonumber\\
\dot{\textbf B}-(1+2\psi)\nabla\times{\textbf E}-2{\textbf E}\times\nabla\phi&=&0
\label{eq:Maxwell}
\end{eqnarray}
where $\rho$ and ${\textbf J}$ are the charge density and electric current respectively and the electric (${\textbf E}$) and magnetic (${\textbf B}$) fields are defined by the relations
\begin{eqnarray}
{\textbf B}&=&-\frac{\nabla\times{\textbf A}}{(1-2\phi)^2}\nonumber\\
{\textbf E}&=&\frac{1}{(1-2\phi)(1+2\psi)}\left(\frac{\partial {\textbf A}}{c\partial t} -\nabla\varphi\right)
\end{eqnarray}
To write down the Klein-Gordon equation 
(Eq. \ref{eq:KGWF}) in its GP form we apply the transformation $\Phi=\Psi e^{-imc^2t/\hbar}$ (see also \cite{Urena-Lopez:2013naa})
\begin{eqnarray}
i(1-2\phi)\hbar \dot\Psi =-\frac{\hbar^2}{2m}\Box_{NEW}\Psi + \frac{\lambda}{2m}|\Psi|^2\Psi&+&\nonumber\\
mc^2(\phi+U_{ext})\Psi &-&\nonumber\\
\left[i\frac{e}{2\hat m}((1+2\phi)\nabla\cdot{\textbf A}-\frac{1}{c}(1-2\psi)\dot\varphi\right.&+&\nonumber\\
\nabla(\psi-\phi)\cdot{\textbf A}+\frac{1}{c}(\dot\psi+3\dot\phi)\varphi)&-&\nonumber\\
\left.\frac{e^2}{2\hat m}((1+2\phi)A^2-(1-2\psi)\varphi^2)\right]\Psi&-&\nonumber\\
i\frac{e}{\hat m}[(1+2\phi){\textbf A}\cdot\nabla\Psi-(1-2\psi)\varphi(\frac{1}{c}\dot\Psi-i\hat m\Psi)].&&
 \label{eq:GP}
\end{eqnarray}
where $\hat m=mc/\hbar$.  This last equation is again the Klein-Gordon equation (Eq. \ref{eq:KG}), but written in terms of the function $\Psi$.

If we remove in Eq. \ref{eq:GP} the gravitational fields $\phi$ and $\psi$ and we set the charge $e=0$, we recover the Gross Pitaevskii equation in the non-relativistic limit. Therefore we interpret Eq. \ref{eq:GP} as the generalization of the GP equation for a relativistic charged bose 
particle in electromagnetic media in a weak gravitational field.
 
In Einstein's theory with a weak gravitational field we have $\phi=\psi$. 
In some alternative theories of gravity this identity does not follow \cite{Aviles:2011ak}.  
For simplicity, we assume a static gravitational
field ($\partial_t\phi=\dot\phi=0$) like that of the Earth. The result after taking the Newtonian limit is
\begin{eqnarray}
i(1-2\phi)\hbar \dot\Psi&=&\nonumber\\
-(1+2\phi)\frac{\hbar^2}{2m}\nabla^2\Psi + \frac{\lambda}{2m}|\Psi|^2\Psi+mc^2\phi\Psi&+&\nonumber\\
mc^2U_{ext}\Psi-i\frac{e}{2\hat m}((1+2\phi)\nabla\cdot{\textbf A}-\frac{1}{c}(1-2\phi)\dot\varphi)\Psi&+&\nonumber\\
\frac{e^2}{2\hat m}((1+2\phi)A^2-(1-2\phi)\varphi^2)\Psi&-& \nonumber\\
i\frac{e}{\hat m}[(1+2\phi){\textbf A}\cdot\nabla\Psi-(1-2\phi)\varphi(\frac{1}{c}\dot\Psi-i\hat m\Psi)].&&
 \label{eq:BECinSchw}
\end{eqnarray}

We interpret Eq. \ref{eq:BECinSchw} as the generalization of the GP 
equation in a gravitational field immersed in a electromagnetic potential. 

We first study the difference between the traditional approach to Newtonian gravity to the 
curved space-time derivation used in this work. For that we neglect the electromagnetic field ($e=0$).
Taking $\phi=0$, we recover the flat space result that corresponds to the usual GP equation
\begin{eqnarray}
i\hbar \dot\Psi =-\frac{\hbar^2}{2m}\nabla^2\Psi + \frac{\lambda}{2m}|\Psi|^2\Psi+
mc^2U_{ext}\Psi
 \label{eq:BECinSchwFlat}
\end{eqnarray}

Taking $U_{ext}=gz/c^2$ 
we obtain the gravitational potential of the Earth ($mgz$). 
Here the gravitational interaction is added as an external potential. This case represents the Newtoninan version of a BEC in a gravitational field.

In the version of Einstein we introduce the gravitational potential 
through the space-time curvature. Taking $U_{ext}=0$ and $\phi=gz/c^2$ 
gives the gravitational potential as well, but there is a correction 
since $\phi$ appears also in the $\dot\Psi$ term of Eq. \ref{eq:BECinSchw}. To determine the size 
of the correction lets consider an atomic interferometer. Here an atom 
(or the BEC wavefunction) is splitted in two, and the two components 
evolve in free fall for some time before recombining them. The separation 
is usually smaller than 1 cm \cite{Dickerson13}, but it is conceivable to 
achieve separations up to 1 m in the near future by combining large free 
fall times with large momentum transfer. A $z=$1 m separation gives a 
$\phi \simeq 10^{-16}$, that is small compared to the 
1 in the $\dot\Psi$ term of Eq. \ref{eq:BECinSchw}.

The combination $gz/c^2$ is the perturbative parameter, in analogy to the term
$v^2/c^2$ of special relativity. The observation of the 
correction requires doing atomic interferometry with 16 digits of precision. 
The current record is at 11 digits \cite{Dickerson13}, but the projected 
sensitivity of the space project is at 15 digits of precision \cite{Geiger11}. 
Having the atoms in space goes in the wrong direction since that eliminates $g$. 
Instead, the present work suggests moving towards atomic interferometry in strong 
gravitational fields.

It is possible to reach higher values of $g$ by taking advantage of the equivalence principle. Instead of the
gravitational field of the Earth one could accelerate the complete experimental setup. This is clearly complicated, but in the case of acceleration introduced by Bloch oscillations things might be simpler. Here it is only necessary to sweep the phase of the two counter propagating lasers beams to produce an accelerated lattice \cite{Clade06}. There is a price to pay in the precision of the measurement due the reduced measurement time since, for example, a linear acceleration of 100 $g$ gives already a displacement of 5 m in 0.1 s. An alternative is to have the BEC in a centrifugal force. Here the gravitational field is $\phi=v^2_c/c^2\ln(R/R_0)$, with $v_c$ the tangential velocity of the BEC, $R$ the radius of the centrifuge and $R_0$ an arbitrary reference radius. For $R=1$ m and $v_c>3.2$ m/s the centrifugal force is bigger than the gravitational force of the Earth.

We apply a Madelung transformation $\Psi=\sqrt{n}e^{iS}$ to Eq \ref{eq:BECinSchw} with $e=0$, with $n$ 
representing the density number of particles and $S$ the velocity super potential 
\begin{eqnarray}
{\textbf v}=\frac{\hbar}{m}\nabla S.
 \label{eq:v}
\end{eqnarray}
The real and imaginary parts of Eq. \ref{eq:BECinSchw} in terms of the $n$ and $S$ variables are
\begin{eqnarray}
(1-2\phi)\dot n+(1+2\phi)\nabla\cdot(n{\textbf v})-(1-2\phi)\dot j&=&0\label{eq:cont}\\
(1-2\phi)\frac{v}{c}+\frac{1}{2}(1+2\phi)\frac{{\textbf v}^2}{c^2}+\phi&+\nonumber\\
+\frac{\lambda}{2m^2}n-\frac{\hbar^2}{2m^2c^2}\frac{\Box_{NEW}\sqrt{n}}{\sqrt{n}}
+\frac{1}{2}(1-2\phi)\frac{v^2}{c^2}&=&0
 \label{eq:Hydro}
\end{eqnarray}
respectively, with the flux ${\textbf j}=n{\textbf v}$ and
\begin{eqnarray}
j=n\frac{\hbar}{mc^2}\dot S=n\frac{v}{c}
 \label{eq:j}
\end{eqnarray}

We interpret Eq. \ref{eq:cont} as the generalized continuity equation in gravitational and electromagnetic fields.  This equation differs from the one derived using pure fluid mechanics in the factors in front of the density and velocity terms. Equations \ref{eq:cont} and \ref{eq:Hydro} are the Klein-Gordon equation in a weak gravitational field, in other words, they are the Einstein-Klein-Gordon equations written in the variables $n$ and $S$. We calculate the gradient of Eq. \ref{eq:Hydro} to obtain the corresponding momentum equation
\begin{eqnarray}
(1-2\phi)n\dot{\textbf v}+(1+2\phi)n({\textbf v}\cdot\nabla){\textbf v}&=&\nonumber\\
n{\textbf F}_{ext}+n(1+2\frac{v}{c}+2\frac{v^2}{c^2}){\textbf F}_{\phi}&-&\nonumber\\
-\nabla p+n{\textbf F}_Q+\nabla\sigma&&
 \label{eq:Momentum}
\end{eqnarray}
where $p$ is the pressure with a state equation $p=(\lambda/m^2) n^2$, ${\textbf F}_Q=-\nabla((\hbar^2/2m^2)\nabla^2\sqrt{n}/\sqrt{n})$ is the quantum force and $\sigma$ is the viscosity (see \cite{Matos:2012qu} for details of the non-gravitational case). As expected, the external force (by unit of mass) is defined by ${\textbf F}_{ext}=-\nabla U_{ext}$ and the gravitational force by ${\textbf F}_{\phi}=-c^2\nabla\phi$. Equations \ref{eq:cont} and \ref{eq:Momentum} correspond to the Klein-Gordon equation (Eq. \ref{eq:KG}) written in terms of $n$ and ${\textbf v}$. 

The difference between the two procedures to include the gravitational potential becomes evident from Eq. \ref{eq:Momentum}. If we consider the gravitational field as an external force (non-covariant case), then space-time is flat, $\phi=0$ and thus ${\textbf F}_{\phi}=0$.  In this case the hydrodynamic equation corresponds to the traditional one. If instead we use the covariant form of the equations we must set ${\textbf F}_{ext}=0$ and include the gravitational force in the ${\textbf F}_{\phi}$ term. The coefficient in front of both terms differs by $2v/c+2v^2/c^2\sim2v/c$. The modification is very small because of the $v$ dependence. To connect with the previous treatment, the wavefunction phase $S$ evolves at a different rate at different heights $z$ because of the gravitational potential, so that $S=mgzt/\hbar$. Using Eq. \ref{eq:j} we have $2v/c=2gz/c^2$ which is the same parameter we obtained before.

In summary, we present a field theoretical approach to describe a BEC in a curved 
space. The derivation results in a generalized GP equation in a gravitational and electromagneitic
fields. We identify the expansion parameter $gz/c^2$ that gives corrections that could be verified with precise enough atomic interferometry. This makes BECs in strong gravitational (or accelerated) 
fields a very interesting candidate to study corrections beyond Newtonian gravity.

\begin{acknowledgements}
This work was partially supported by CONACyT M\'exico under grants
CB-2009-01, no. 132400, CB-2011, no. 166212, Xiuhcoatl cluster at Cinvestav 
and I0101/131/07
C-234/07 of the Instituto Avanzado de Cosmolog\'ia (IAC)
collaboration (http://www.iac.edu.mx/). EG aknowledges support from CONACyT and Fundaci\'on Marcos Moshinsky
\end{acknowledgements}

\end{document}